\documentclass[prl,aps,twocolumn,floatfix,showpacs]{revtex4}
\usepackage{graphics}
\begin{document}
\title{Evolution from BCS to BKT superfluidity in one-dimensional optical lattices}
\author{M. Iskin$^1$ and C. A. R. S{\'a} de Melo$^{1,2}$}
\affiliation{$^1$Joint Quantum Institute, National Institute of Standards and Technology 
and University of Maryland, Gaithersburg, Maryland 20899-8423, USA. \\
$^2$School of Physics, Georgia Institute of Technology, Atlanta, Georgia 30332-0432, USA.}
\date{\today}

\begin{abstract}

We analyze the finite temperature phase diagram of fermion mixtures in one-dimensional 
optical lattices as a function of interaction strength. At low temperatures, 
the system evolves from an anisotropic three-dimensional Bardeen-Cooper-Schrieffer (BCS)
superfluid to an effectively two-dimensional Berezinskii-Kosterlitz-Thouless (BKT)
superfluid as the interaction strength increases. We calculate the critical temperature 
as a function of interaction strength, and identify the region where the dimensional 
crossover occurs for a specified optical lattice potential. Finally, we show that the 
dominant vortex excitations near the critical temperature evolve from multiplane elliptical 
vortex loops in the three-dimensional regime to planar vortex-antivortex pairs in the 
two-dimensional regime, and we propose a detection scheme for these excitations.

\end{abstract}
\maketitle

Ultracold atoms in optical lattices are ideal systems to simulate and study 
novel and exotic condensed matter phases. Remarkable success has been 
achieved experimentally with Bose atoms loaded into three-dimensional (3D) 
optical lattices, where superfluid and Mott-insulator phases have been 
observed~\cite{bloch-2005}. In addition, experimental evidence for superfluid
and possibly insulating phases were found for fermionic atoms ($^6$Li) in 3D optical 
lattices~\cite{ketterle-2006}. Compared with the purely homogeneous 
or harmonically trapped systems, optical lattices offer additional flexibilities 
and an unprecedented degree of control such that their physical properties 
can be studied as a function of onsite atom-atom interactions, tunneling 
amplitudes between adjacent sites, atom filling fractions and lattice dimensionality.

For instance, in strictly two-dimensional (2D) systems the superfluid transition
for bosons and fermions is of the Berezinskii-Kosterlitz-Thouless (BKT) 
type~\cite{berezinskii-1971,kosterlitz-1972}. This phase is characterized by the 
existence of bound vortex-antivortex pairs below the critical temperature 
$T_{BKT}$, and evidence for it was recently reported in nearly 2D 
Bose gases confined to one-dimensional (1D) optical lattices~\cite{zoran-2006}.
Thus, it is very likely that one of the next research frontiers for experiments 
with fermions in optical lattices is also the investigation of such a transition.

For bosons or fermions, it is possible to study not only 3D and 2D superfluids 
as two separate limits, but also the entire evolution from 3D to 2D by tuning nearly 
continuously the tunneling amplitudes~\cite{stoferle-2006, bongs-2006}.
However, fermions offer the additional advantage that their interactions can also
be tuned using Feshbach resonances without having to worry about the collapse
of the condensate, as it is the case for bosons. Furthermore, the phase diagram
of fermions in optical lattices also shows superfluid-to-insulator 
transitions~\cite{iskin-2007, ho-2007, sachdev-2007} like bosons do.

Anticipating experiments, we study in this manuscript the finite 
temperature phase diagram of attractive fermion mixtures in 1D optical lattices,
and discuss the dimensional crossover from an anisotropic-3D BCS superfluid to 
an effectively 2D BKT superfluid as a function of interaction strength and tunneling 
parameters. We show that vortex excitations near the critical temperature
change from elliptical multiplane vortex loops in the anisotropic-3D BCS regime to
planar vortex-antivortex pairs in the 2D BKT regime. 
Finally, we propose an experiment for the detection of vortex excitations.

%
%
To describe fermion mixtures in 1D optical lattices, we start with the Hamiltonian ($\hbar = k_B = 1$)
\begin{equation}
H = \sum_{\mathbf{k},\sigma} \xi_{\mathbf{k},\sigma} a_{\mathbf{k},\sigma}^\dagger a_{\mathbf{k},\sigma} 
- g\sum_{\mathbf{k},\mathbf{k'},\mathbf{q}} \Gamma_{\mathbf{k}}^* \Gamma_{\mathbf{k'}} 
b_{\mathbf{k},\mathbf{q}}^\dagger b_{\mathbf{k'},\mathbf{q}}, 
\label{eqn:hamiltonian}
\end{equation}
where the operator $ a_{\mathbf{k},\sigma}^\dagger$ creates a fermion with pseudo-spin 
$\sigma$ which labels either the type of atoms for unequal mass mixtures or the 
hyperfine state of atoms for equal mass mixtures. The operator
$
b_{\mathbf{k},\mathbf{q}}^\dagger = a_{\mathbf{k}+\mathbf{q}/2,\uparrow}^\dagger 
a_{-\mathbf{k}+\mathbf{q}/2,\downarrow}^\dagger
$ 
creates fermion pairs with center of mass momentum $\mathbf{q}$ and relative momentum $2\mathbf{k}$,
while $g > 0$ and $\Gamma_{\mathbf{k}}$ are the strength and symmetry of the attractive 
interaction between fermions, respectively. Here, 
$
\xi_{\mathbf{k},\sigma}= \epsilon_{\mathbf{k},\sigma} - \mu_\sigma
$ 
represents the difference between the kinetic energy
\begin{equation}
\epsilon_{\mathbf{k},\sigma} = k_\perp^2 / (2m_{\sigma}) + 2t_{z,\sigma}\left[1 - \cos(k_z a_z)\right]
\end{equation}
and the chemical potential $\mu_\sigma$, and $a_z$ is the lattice spacing along the 
$\mathbf{z}$ direction. We allow for the fermions to have different masses $m_{\sigma}$ 
and tunneling amplitudes $t_{z,\sigma}$, but we confine our analysis to equal 
population mixtures.

The saddle-point action for this Hamiltonian is
\begin{eqnarray}
S_0(\Delta_0^*, \Delta_0) &=& \beta |\Delta_0|^2/g +
(1/M)\sum_\mathbf{k}\big\lbrace
\beta(\xi_{\mathbf{k},\downarrow} - E_{\mathbf{k},2}) \nonumber \\
&+& 
\ln [(1+{\cal X}_{\mathbf{k},1})/2] + \ln [(1+{\cal X}_{\mathbf{k},2})/2]
\big\rbrace, 
\end{eqnarray}
where $\beta = 1/T$ is the inverse temperature, $M$ is the number of lattice 
sites along the $\mathbf{z}$ direction,
$
E_{\mathbf{k},s} = (\xi_{\mathbf{k},+}^2 + 
|\Delta_{\mathbf{k}}|^2)^{1/2} + \gamma_s \xi_{\mathbf{k},-}
$
is the quasiparticle energy when $\gamma_1 = 1$ and the negative of the quasihole 
energy when $\gamma_2 = -1$.
Here, 
$
\xi_{\mathbf{k},\pm} = (\xi_{\mathbf{k},\uparrow} \pm \xi_{\mathbf{k},\downarrow})/2,
$
$
{\cal X}_{\mathbf{k},s} = \tanh(\beta E_{\mathbf{k},s}/2),
$
and $\Delta_{\mathbf{k}} = \Delta_0 \Gamma_{\mathbf{k}}$ is the saddle-point order parameter.

%
%
The order parameter equation is obtained from the stationary condition 
$
\partial S_0 / \partial \Delta_0^* = 0,
$ 
leading to
\begin{equation}
1/g = (1/M) \sum_{\mathbf{k}} 
|\Gamma_{\mathbf{k}}|^2 {\cal X}_{\mathbf{k},+} / (2E_{\mathbf{k},+}),
\label{eqn:sp.op}
\end{equation}
where 
$
{\cal X}_{\mathbf{k},\pm} = ({\cal X}_{\mathbf{k},1} \pm {\cal X}_{\mathbf{k},2})/2
$
and
$
E_{\mathbf{k},\pm} = (E_{\mathbf{k},1} \pm E_{\mathbf{k},2})/2.
$
We may eliminate $g$ in favor of the binding energy $\epsilon_b < 0$ of 
two fermions in the lattice potential via 
$
1/g = (1/M) \sum_{\mathbf{k}} |\Gamma_{\mathbf{k}}|^2 / 
(\epsilon_{\mathbf{k},\uparrow} + \epsilon_{\mathbf{k},\downarrow} - \epsilon_b).
$
For s-wave interactions with range $R_0 \sim k_0^{-1}$, we take $\Gamma_\mathbf{k} = 1$ 
for $k < k_0$ and zero otherwise, leading to
\begin{equation}
\epsilon_b  = 4t_{z,+} - (2t_{z,+}^2/\epsilon_0) \exp(1/G) - 2\epsilon_0 \exp(-1/G),
\end{equation}
where $A$ is the area in the $(x,y)$ plane, $\epsilon_0 = k_0^2/(2m_+)$,
$
t_{z,+} = (t_{z,\uparrow} + t_{z,\downarrow})/2,
$ 
$G = m_+ A g /(4\pi)$ is the dimensionless interaction strength, and
$
m_\pm = 2m_\uparrow m_\downarrow/(m_\uparrow \pm m_\downarrow).
$ 
Notice that two-body bound states in vacuum only exist beyond a critical interaction strength
$
G_c = 1/\ln(\epsilon_0/t_{z,+})
$
for finite $t_{z,+}$, while they always exist for arbitrarilly small $G$ 
in the 2D limit where $t_{z,+} \to 0$.

Eq.~(\ref{eqn:sp.op}) has to be solved self-consistently with the number equation 
$
N_{0, \sigma} = -\partial S_0 / (\beta \partial \mu_\sigma),
$ 
leading to
\begin{equation}
N_{0,\sigma} = \sum_{\mathbf{k}} 
\left[ (1 - \gamma_s {\cal X}_{\mathbf{k},-})/2
- \xi_{\mathbf{k},+} {\cal X}_{\mathbf{k},+} / (2E_{\mathbf{k},+})
\right].
\label{eqn:sp.number}
\end{equation}
Solutions to Eqs.~(\ref{eqn:sp.op}) and~(\ref{eqn:sp.number}) constitute an 
approximate description of the system only when amplitude and phase fluctuations 
of the order parameter are small, which is the case only at low temperatures, although
quantum fluctuations play a role. However, fluctuations are extremely important
close to the critical temperature $T_c$.

%
%
The derivation of the fluctuation action is accomplished by writing the order parameter as
$
\Phi(q) = |\Delta_0| \delta_{q,0}+ \lambda(q)
$
with $\lambda(q) = |\lambda(q)| e^{i\theta(q)}$, where $|\lambda(q)|$ is the amplitude 
and $\theta(q)$ is the phase of the fluctuations.
Near $T_c$, $|\Delta_0|$ vanishes, and the fluctuation action reduces to
$
S_{fl}(\lambda^*,\lambda) = \beta \sum_{q} \lambda^*(q) L^{-1}(q) \lambda(q)
+ (\beta b/2) \sum_{q} |\lambda(q)|^4,
$
where the quadratic term is 
\begin{eqnarray}
L^{-1}(q) = \frac{1}{g} - \frac{1}{2M}\sum_{\mathbf{k}} 
\frac{X_{\mathbf{q}/2 + \mathbf{k},\uparrow} + X_{\mathbf{q}/2 - \mathbf{k},\downarrow}} 
{\xi_{\mathbf{q}/2 + \mathbf{k},\uparrow} + 
\xi_{\mathbf{q}/2 - \mathbf{k},\downarrow} - iv_\ell} |\Gamma_{\mathbf{k}}|^2,
\label{eqn:L-1}
\end{eqnarray}
and the quartic term is 
$
b = \sum_\mathbf{k} [ X_{\mathbf{k},+} / (4 \xi_{\mathbf{k},+}^3) 
- \beta Y_{\mathbf{k},+}/(8 \xi_{\mathbf{k},+}^2)].
$
Here, 
$X_{\mathbf{k},\sigma} = \tanh (\beta \xi_{\mathbf{k},\sigma}/2)$,
$Y_{\mathbf{k},\sigma} = {\rm sech}^2 (\beta \xi_{\mathbf{k},\sigma}/2)$
are thermal factors, and
$X_{\mathbf{k},\pm} = (X_{\mathbf{k},\uparrow} \pm X_{\mathbf{k},\downarrow})/2$ and 
$Y_{\mathbf{k},\pm} = (Y_{\mathbf{k},\uparrow} \pm Y_{\mathbf{k},\downarrow})/2$.

The analytic continuation $i v_{\ell} \to \omega + i\delta$ where $\delta \to 0$, and a long wavelength 
and low frequency expansion leads to 
$
L^{-1} (q) = a + \sum_{ij} q_i c_{i,j} q_j - d \omega.
$
The momentum and frequency independent coefficient is
$
a = 1/g - (1/M)\sum_{\mathbf{k}} X_{\mathbf{k},+}
|\Gamma_\mathbf{k}|^2 / (2\xi_{\mathbf{k},+});
$
the coefficients for low momentum are 
$
c_{ij} = \sum_{\mathbf{k}} \big\lbrace
\beta^2
	( \dot{\xi}_{\mathbf{k},\uparrow}^i \dot{\xi}_{\mathbf{k},\uparrow}^j X_{\mathbf{k},\uparrow} Y_{\mathbf{k},\uparrow} 
	+ \dot{\xi}_{\mathbf{k},\downarrow}^i \dot{\xi}_{\mathbf{k},\downarrow}^j X_{\mathbf{k},\downarrow} Y_{\mathbf{k},\downarrow})
/ (16\xi_{\mathbf{k},+})
+ \beta
	[ (\dot{\xi}_{\mathbf{k},\uparrow}^i-\dot{\xi}_{\mathbf{k},\downarrow}^i)
	  (\dot{\xi}_{\mathbf{k},\uparrow}^j Y_{\mathbf{k},\uparrow} - \dot{\xi}_{\mathbf{k},\downarrow}^j Y_{\mathbf{k},\downarrow}) 
	- \xi_{\mathbf{k},+} (\ddot{\xi}_{\mathbf{k},\uparrow}^{i,j} Y_{\mathbf{k},\uparrow} + \ddot{\xi}_{\mathbf{k},\downarrow}^{i,j} Y_{\mathbf{k},\downarrow})]
/ (16\xi_{\mathbf{k},+}^2)
+ X_{\mathbf{k},+} [\xi_{\mathbf{k},+} (\ddot{\xi}_{\mathbf{k},\uparrow}^{i,j} + \ddot{\xi}_{\mathbf{k},\downarrow}^{i,j})
										-(\dot{\xi}_{\mathbf{k},\uparrow}^i - \dot{\xi}_{\mathbf{k},\downarrow}^i)
										 (\dot{\xi}_{\mathbf{k},\uparrow}^j - \dot{\xi}_{\mathbf{k},\downarrow}^j)
] / (8\xi_{\mathbf{k},+}^3)
\big\rbrace |\Gamma_\mathbf{k}|^2
,
$
where $\dot{\xi}_{\mathbf{k},\sigma}^i = \partial \xi_{\mathbf{k},\sigma}/\partial k_i$ and
$\ddot{\xi}_{\mathbf{k},\sigma}^{i,j} = \partial^2 \xi_{\mathbf{k},\sigma}/(\partial k_i \partial k_j)$;
and finally the coefficient for low frequency is 
$
d = \lim_{\omega \to 0} \sum_{\mathbf{k}}
X_{\mathbf{k},+} 
[ 1/(4 \xi_{\mathbf{k},+}^2) + i \pi \delta(2\xi_{\mathbf{k},+} - \omega)/\omega
] |\Gamma_\mathbf{k}|^2.
$
Notice that, $c_{ij}$ is diagonal for s-wave symmetry with 
$c_{ij} = c_i \delta_{ij}$ and $c_\perp \equiv c_x = c_y \ne c_z$ 
leading to
$
L^{-1} (q) = a + c_\perp q_\perp^2 + c_z q_z^2 - d \omega.
$
Here, $\delta_{ij}$ is the Kronecker-delta.

%
%
We consider first the strong attraction regime ($G \gg 1$) corresponding to $\mu_\sigma < 0$ 
and $|\mu_\sigma| \sim |\epsilon_b|/2 \gg t_{z,+}$, where
$
L^{-1}(q) = m_+ A /(4\pi |\epsilon_b|) [iv_\ell - \omega_B (\mathbf{q}) + 2\mu_B]
$
to lowest order of $\mathbf{q}$ and $v_\ell$, with 
\begin{equation}
\omega_B (\mathbf{q}) = q_\perp^2/(2m_{B,\perp}) + 2t_{B,z} \left[1 - \cos(q_z a_z)\right]. 
\end{equation}
After the rescaling $\Psi(q) =  \sqrt{ m_+ A /(4\pi |\epsilon_b| )} \lambda(q)$, 
the quadratic term of $S_{fl}$ describes non-interacting bosons with dispersion 
$\omega_B (\mathbf{q})$, mass $m_{B,\perp} = m_\uparrow + m_\downarrow$ in the 
$(x,y)$ plane, tunneling amplitude $t_{B,z} = 2t_{z,\uparrow} t_{z,\downarrow} / |\epsilon_b|$
along the $\mathbf{z}$ direction, and chemical potential $\mu_B = \mu_\uparrow + \mu_\downarrow - \epsilon_b$.
Since the quartic term of $S_{fl}$ is small, the resulting Bose gas is weakly 
interacting, leading to a dominant contribution to the number equation
\begin{equation}
\label{eqn:number-fluctuation}
N_{fl, \sigma} = \sum_{\mathbf{q}} n_B [\omega_B (\mathbf{q}) - \widetilde{\mu}_B ],
\end{equation}
which is the same for $\uparrow$ and $\downarrow$ fermions.
Here, $n_B(x) = 1/(e^{\beta x} - 1)$ is the Bose distribution 
and $\widetilde{\mu}_B = \mu_B - V_H < 0$ includes the Hartree shift $V_H$.

For $G \gg 1$, Eq.~(\ref{eqn:number-fluctuation}) leads to  Bose-Einstein condensation 
of tightly bound fermion pairs at $\mathbf{q} = \mathbf{0}$ with
$
T_c  = (m_+/m_{B, \perp})\epsilon_{2D} / \ln(T_c/t_{B,z}),
$
where $\epsilon_{2D} = k_{2D}^2/(2m_+)$ is a characteristic energy of fermions in 2D. Here, 
$k_{2D}$ is a 2D momentum defined through the 2D density 
$n_{2D} = k_{2D}^2/(2\pi)$. We also define an effective 3D density $n_{3D} = n_{2D}/a_z$ 
where $n_{3D} = k_{3D}^3/(3\pi^2)$, and $k_{3D}$ is the 3D momentum. 
Notice that, 
$
\epsilon_{2D} = [2 k_{3D} a_z /(3\pi)] \epsilon_{3D}
$ 
where $\epsilon_{3D} = k_{3D}^2/(2 m_+)$ is a characteristic energy in 3D.

\begin{figure} [htb]
\centerline{\scalebox{0.5}{\includegraphics{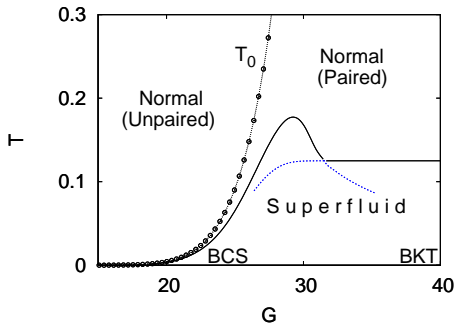}}}
\caption{\label{fig:BEC}
Phase diagram of temperature $T$ (in units of $\epsilon_{2D}$) versus
$G = mAg/(4\pi)$ for a mixture of equal masses with s-wave interactions, 
interaction range $k_0 \sim 10^4k_{2D}$, tunneling $t_z \approx 0.043\epsilon_{2D}$,
lattice spacing $a_z \approx 0.43{\rm \mu m}$ and planar density 
$n_{2D} \approx 2.5 \times 10^7 {\rm cm^{-2}}$, such that $k_{2D} a_z \approx 2.2$.
$T_0$ is the saddle-point or pairing temperature scale.
}
\end{figure}

For fixed $t_{z,\sigma}$, Eq.~(\ref{eqn:number-fluctuation}) shows that $T_c$ is a decreasing function 
of $G$. This is most easily seen for a dilute system where 
$2t_{B,z} [1 - \cos(q_z a_z)] \approx q_z^2/(2m_{B,z})$,
such that $m_{B,z} = 1/(2t_{B,z} a_z^2)$ is the effective mass along the $\mathbf{z}$ direction. 
In this case, Eq.~(\ref{eqn:number-fluctuation}) gives
$
T_c  \approx 0.218 [2m_+/(m_{B,\perp}^2 m_{B,z})^{1/3}] \epsilon_{3D},
$
which reduces to the 3D continuum result $T_c = 0.218 \epsilon_F$ of equal mass 
mixtures~\cite{sademelo-1993} where $m_+ = m$, $m_{B,\perp} = m_{B,z} = 2m$ and 
$\epsilon_{3D} \equiv \epsilon_F$.
However, $T_c \to 0$ asymptotically when $t_{B,z} \to 0$ or $m_{B,z} \to \infty$,
which occurs when the binding energy becomes very large 
$(|\epsilon_b| \gg \sqrt{t_{z,\uparrow} t_{z,\downarrow}})$.
This limit is clearly unphysical and shows the breakdown of the Gaussian theory 
in 1D optical lattices, since the BKT transition of tightly bound fermion pairs is not 
recovered in the 2D limit, as can be seen in Fig.~\ref{fig:BEC}.

%

%
%
To recover the BKT physics in the $G \gg 1$ limit where the paired fermions live 
in 2D planes, we return to the derivation of the fluctuation action $S_{fl}$ 
with $t_{z, \sigma} = 0$. Taking the order parameter as
$
\Phi(x) = [|\Delta_0| + |\eta(x)|] e^{i\theta(x)},
$
where $|\eta(x)|$ corresponds to the amplitude fluctuations and $\theta(x)$ is the 
phase of the order parameter such that $|\Delta_0| \gg |\eta(x)|$, 
we obtain the phase-only action
$
S_{fl}(\theta) = (\beta/2) \sum_{q} \left[ \kappa_0(T) v_\ell^2 
+ q_i \rho_{i,j}(T) q_j \right] \theta(q)\theta(-q).
$
Here, the coefficient 
$
\kappa_0(T) = (1/4)\sum_{\mathbf{k}} [ 
|\Delta_\mathbf{k}|^2 {\cal X}_{\mathbf{k},+}/E_{\mathbf{k},+}^3 
+ \beta \xi_{\mathbf{k},+}^2 {\cal Y}_{\mathbf{k},+}/(2E_{\mathbf{k},+}^2) ]
$
is the atomic compressibility, where
$
{\cal Y}_{\mathbf{k},\pm} = ({\cal Y}_{\mathbf{k},1} \pm {\cal Y}_{\mathbf{k},2})/2
$
with ${\cal Y}_{\mathbf{k},s} = {\rm sech}^2(\beta E_{\mathbf{k},s}/2)$;
while the coefficient of the gradient term
\begin{eqnarray}
\rho_{ij}(T) &=& \frac{n \delta_{ij}}{4m_+} - 
\sum_{\mathbf{k}} \frac{k_i k_j}{16A} \bigg\lbrace 
\frac{4|\Delta_{\mathbf{k}}|^2}{m_-^2}
\left( \frac{{\cal X}_{\mathbf{k},+}}{E_{\mathbf{k},+}^3} - 
\frac{\beta {\cal Y}_{\mathbf{k},+}}{2 E_{\mathbf{k},+}^2} \right) \nonumber \\
&+&
\beta\left[ \frac{{\cal Y}_{\mathbf{k},1}}{m_\uparrow^2}
+ \frac{{\cal Y}_{\mathbf{k},2}}{m_\downarrow^2}
- \frac{4 {\cal Y}_{\mathbf{k},-}}{m_+ m_-}
\left( 1 - \frac{\xi_{\mathbf{k},+}}{E_{\mathbf{k},+}}\right)
\right]
\bigg\rbrace
\end{eqnarray}
is the phase stiffness, where $\rho_{ij}(T) = \rho_{0}(T) \delta_{ij}$ for the s-wave symmetry.

This leads to the BKT transition temperature~\cite{berezinskii-1971,kosterlitz-1972}
\begin{equation}
T_{BKT}  = \pi \rho_0(T_{BKT}) / 2,
\label{eqn:bkt}
\end{equation}
which needs to be solved self-consistently with Eqs.~(\ref{eqn:sp.op}) and~(\ref{eqn:sp.number}) 
in order to determine $T_{BKT}$, $|\Delta_0|$ and $\mu_\sigma$ as a function of $G$.
In the weak attraction regime ($G \lesssim 1$), $T_{BKT}$ increases with $G$ as
$
T_{BKT} = (e^\gamma/\pi) (2|\epsilon_b| \epsilon_{2D})^{1/2},
$
where $\gamma \approx 0.577$ is the Euler's constant and
$
\epsilon_b = -2\epsilon_0\exp(-1/G)
$
is the binding energy in 2D. While in the strong attraction regime ($G \gg 1$), $T_{BKT}$ saturates to
$
T_{BKT} = (\epsilon_{2D}/8) [1 - 2m_+^2/(3m_-^2)].
$
For equal mass mixtures ($m_{-} \to \infty$), this reduces to $T_{BKT} = 0.125\epsilon_F$ 
which can be seen in Fig.~\ref{fig:BEC}. Here, $\epsilon_F \equiv \epsilon_{2D}$ is the 2D Fermi energy.

%
%
To estimate when $G$ induces the crossover from anisotropic-3D to 2D behavior, we compare 
the critical temperature $T_{c, Gauss}$ obtained from the Gaussian theory with the 
critical temperature $T_{BKT}$ for the BKT transition in the strict 2D limit. 
When $G \gg 1$, the condition $T_{c, Gauss} = T_{BKT}$ leads to
\begin{equation}
\label{eqn:crossover-tc}
t_{z,c,\uparrow} t_{z,c,\downarrow}= 
\frac{\zeta^2(3/2)}{ 2^{12}\pi } 
\left( \frac{m_B}{m_+} \right)^2 
\left( 1 - \frac{2m_+^2}{3m_-^2} \right)^3
\epsilon_{2D}|\epsilon_b|,
\end{equation}
where $\zeta(x)$ is the zeta function. This relation reduces to 
$
t_{z,c} = \left[ \zeta(3/2)/(32 \sqrt{\pi}) \right] (\epsilon_{2D} |\epsilon_b|)^{1/2}
$
for mixtures of equal mass fermions and equal tunneling.

\begin{figure} [htb]
\centerline{\scalebox{0.38}{\hskip 15mm \includegraphics{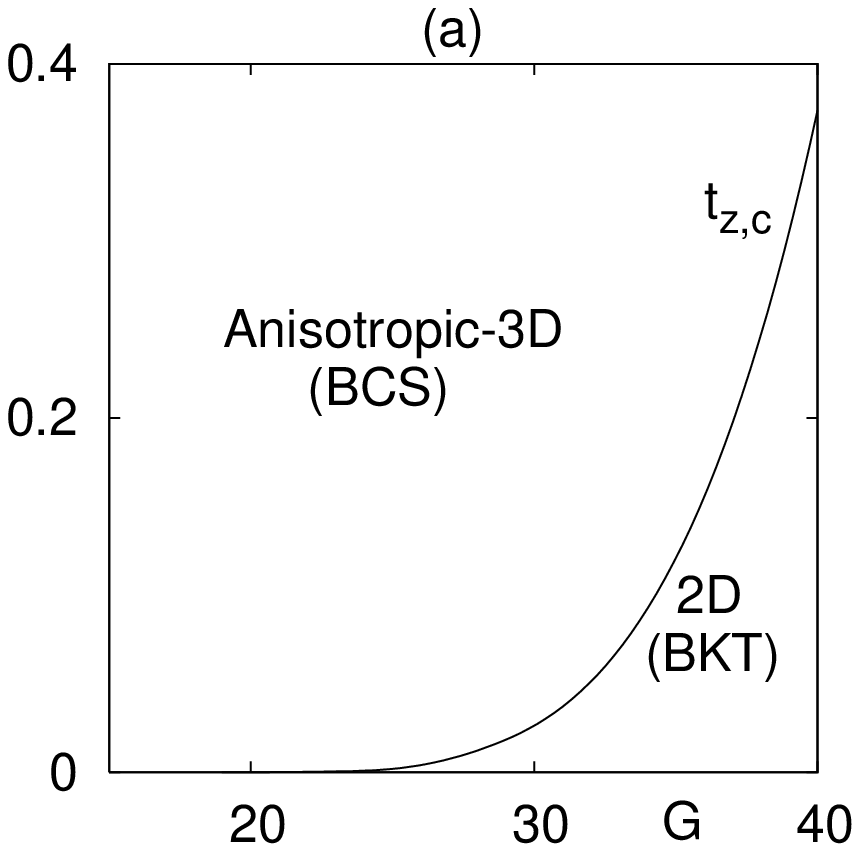} 
\hskip -25mm \includegraphics{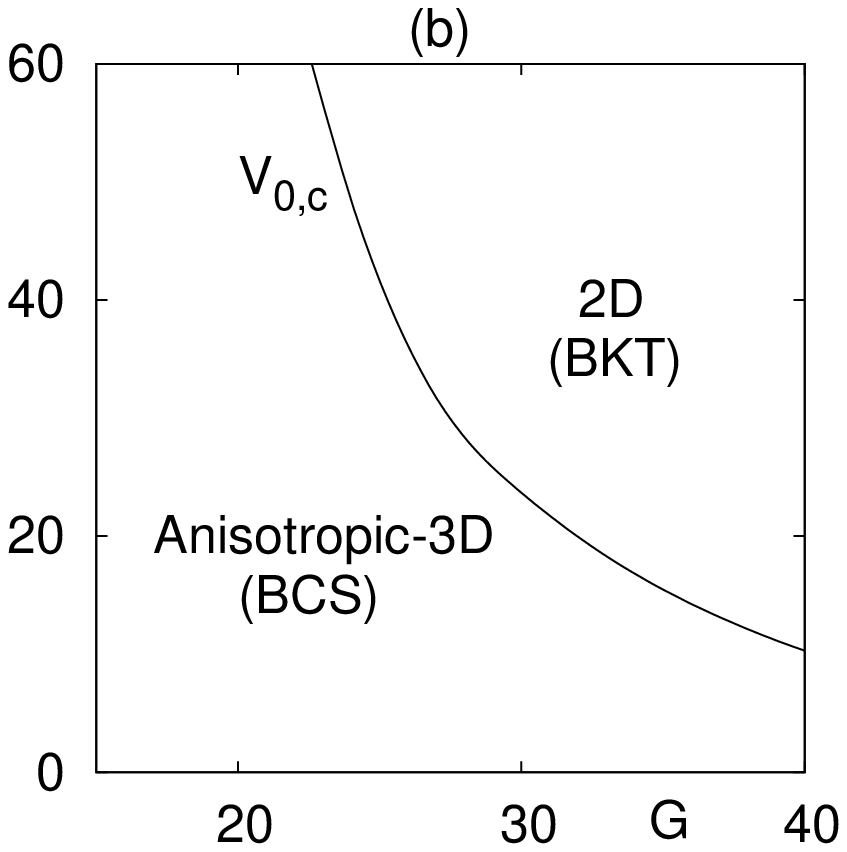}}}
\caption{\label{fig:QUASI}
Characteristic (a) tunneling amplitude $t_{z,c}$ (in units of $\epsilon_{2D}$); and
(b) optical lattice depth $V_{0,c}$ (in units of $E_r$) versus $G = mAg/(4\pi)$ showing 
the anisotropic-3D to 2D crossover. The parameters are the same as in Fig.~\ref{fig:BEC}.
}
\end{figure}

We can also relate $t_{z,\sigma}$ to the depth $V_{0,\sigma}$ of the 1D optical latttice 
potential $V_\sigma (\mathbf{r}) = V_{0,\sigma} \sin^2 (\pi z/a_z)$ where
$
t_{z,\sigma} = (2 E_{r,\sigma}/\sqrt{\pi}) (\alpha_{\sigma})^{3/4} \exp (- 2 \sqrt{\alpha_{\sigma}}).
$
Here, $\alpha_\sigma = V_{0,\sigma}/E_{r,\sigma}$ where $E_{r,\sigma} = \pi^2 / (2 m_\sigma a_z^2)$ 
is the recoil energy. 
In Fig.~\ref{fig:QUASI}, we show the characteristic $t_{z,c} = t_{z,c,\sigma}$ and
$V_{0,c} = V_{0,c,\sigma}$ lines which separate the anisotropic-3D from the 2D regime,
for mixtures of equal mass, equal tunneling and s-wave interactions. 
When $G$ is fixed, the 2D regime may be reached from the anisotropic-3D regime with 
increasing $V_0 = V_{0,\sigma}$ or decreasing $t_z = t_{z,\sigma}$. While for fixed 
$V_0 = V_{0,\sigma}$ or $t_z = t_{z,\sigma}$, the 2D regime may be reached from the 
anisotropic-3D regime by increasing $G$.

%

%
%
Further insight into the dimensional crossover is gained by 
rewriting $S_{fl}$ with $t_{z, \sigma} \ne 0$ 
in real space and time such that
$
S_{fl}(\lambda^*, \lambda) =  (1/V) \int dt dr_\perp d z {\cal L}_{fl}(\lambda^*, \lambda),
$
where
$
{\cal L}_{fl}(\lambda^*, \lambda) =  \lambda^*(x) 
({\cal O} - c_{z} \partial_{z}^{2}) \lambda(x) + b |\lambda(x)|^4/2 
$
is the Lagrangian. Here, $V$ is the volume, $\lambda(x) \equiv \lambda (r_\perp, z, t)$ 
is the fluctuation field and 
$
{\cal O} = a - c_\perp \nabla_\perp^2 - i d \partial_t.
$
Upon discretization $z = n a_z$, $S_{fl}$ reduces to the Lawrence-Doniach (LD) action 
$
S_{LD}(\lambda_n^*, \lambda_n) = [1/(MA)] \sum_n \int dt dr_\perp {\cal L}_{LD}(\lambda_n^*, \lambda_n),
$
where
\begin{eqnarray}
\label{eqn:LD-action}
{\cal L}_{LD}(\lambda_n^*, \lambda_n) = \lambda_n^* {\cal O} \lambda_n 
+ \frac{c_z}{a_z^2} |\lambda_{n+1} - \lambda_n|^2 + \frac{b}{2}|\lambda_n|^4
\end{eqnarray}
is the LD Lagrangian~\cite{lawrence-doniach-1971}. Here, the local field 
$
\lambda_n \equiv \lambda (r_\perp, z = n a_z, t)
$
describes the order parameter in each plane labeled by index $n$. Writing $a = a_0 \epsilon (T)$ with 
$
\epsilon (T) = (T - T_c)/T_c,
$
scaling the field $\psi_n = \sqrt{b/a_0}\lambda_n$, and defining the correlation lengths 
$\xi_{0,\perp}^2 = c_\perp/a_0$ and $\xi_{0,z}^2 = c_z/a_0$, and the characteristic time 
$\tau_0 = -d/a_0$ leads to the scaled action
$
\widetilde{\cal L}_{LD}(\psi_n^*, \psi_n) = \psi_n^*  i \tau_0 \partial_t  \psi_n 
+ \epsilon(T) |\psi_n|^2 + \xi_{0,\perp}^2 |\nabla_\perp\psi_n|^2 
+ \xi_{0,z}^2 |\psi_{n+1}-\psi_n|^2/a_z^2 + |\psi_n|^4/2
$,
which describes the system near $T_c$. Here, 
$
\widetilde{\cal L}_{LD}(\psi_n^*, \psi_n) = (b/a_0^2) {\cal L}_{LD}(\lambda_n^*, \lambda_n).
$ 
Furthermore, taking $\psi_n = |\psi_n|\exp (i\theta_n)$ in the LD action, such
that $|\psi_n| = \phi_0$ is independent of position and time, leads to the 
phase-only anisotropic-3D XY model with the dimensionless Hamiltonian 
\begin{eqnarray*}
\widetilde{H}_{XY} (r_\perp, n) = K_\perp |a_\perp \nabla_\perp \theta_n|^2  
- 2 K_z \cos (\theta_{n + 1} - \theta_n) + C, 
\end{eqnarray*}
where $K_\perp = (\xi_{0,\perp} \phi_0/a_\perp)^2$, $K_z = (\xi_{0,z} \phi_0/a_z)^2$ 
and $a_\perp \sim k_{2D}^{-1}$, and $C$ is a constant. The dimension-full Hamiltonian
is 
$
\widetilde{H}_{XY} (r_\perp, n) = (b/a_0^2) H_{XY} (r_\perp, n).
$

This can be mapped into the vortex-loop representation~\cite{shenoy-1995} yielding the dual 
dimensionless Hamiltonian
\begin{eqnarray*}
\widetilde{H}_D =
\pi \sum_{\mathbf{r} \ne \mathbf{r'}} 
[ K_z J_\perp (\mathbf{r}) \cdot J_\perp (\mathbf{r'}) 
+ K_\perp J_z (\mathbf{r}) \cdot J_z (\mathbf{r'}) ] U (\mathbf{R}),
\end{eqnarray*}
where $U (\mathbf{R} = \mathbf{r} - \mathbf{r'})$ plays the role of an interaction potential 
for the vortex loop field 
$
\mathbf{J} (\mathbf{r}) = [ J_\perp (\mathbf{r}), J_z (\mathbf{r})],
$
and satisfies the differential equation 
$
(\nabla_\perp^2 + \eta^{-2} \partial_z^2 ) U(\mathbf{R}) = - 4 \pi \delta (\mathbf{R}).
$
Here, $\eta = \sqrt{K_\perp/ K_z}$ is the anisotropy ratio and $\delta(x)$ is the delta function.
The dual transformation maps closed supercurrent flows associated with the gradients of the 
phase $\theta$ into the vortex-loop vector $\mathbf{J} (\mathbf{r})$, in the same way that 
the electric current flowing on a ring can be mapped into a magnetic field vector 
with the help of the Biot-Savart law.

For large magnitudes of $\mathbf{R} = (R_\perp, R_z)$, the vortex-loop interaction behaves 
asymptotically as 
$
U (\mathbf{R}) \sim 1/\sqrt{ [R_\perp/ (\eta a_\perp)]^2 + (R_z/a_z)^2 },
$ 
and leads to equipotentials in the shape of ellipsoids 
$
[R_\perp/(\eta a_\perp)]^2 + (R_z/a_z)^2  = U_0^{-2}
$
when $U (\mathbf{R}) = U_0$. Elliptical vortex loops corresponding to a nearly toroidal 
arrangement of the supercurrent flow are the large scale excitations formed by a continuous 
closed line having the same potential between segments with $\mathbf{r}= - \mathbf{r'}$.
When $\eta \to \infty$, the planes along the $\mathbf{z}$ direction decouple (2D BKT regime)
and the vortex loops reduce to planar vortex-antivortex pairs. For $2 < \eta < \infty$, 
the system is still nearly 2D, and the dominant excitations are square vortex loops coupling 
two consecutive planes and planar vortex loops.
However, in the anisotropic-3D regime when $\eta < 2$, the dominant excitations become 
multiplane elliptical vortex loops.

In the strong attraction regime $(G \gg 1)$,
$
\eta \approx (a_z/ a_{\perp}) \sqrt{m_{B,z}/m_{B,\perp}} \gg 1,
$
and the 2D BKT limit is recovered since 
$m_{B,z} \gg m_{B,\perp}$.
For fermion mixtures of equal masses and equal tunnelings, we can rewrite this condition as
$
\eta \approx \sqrt{\epsilon_{2D} |\epsilon_b|}/(4\pi t_z) \gg 1. 
$
Since the anisotropic-3D to 2D crossover occurs for $\eta \approx 2$, 
this condition leads to $t_{z,c} \approx (1/8\pi) (\epsilon_{2D} \vert \epsilon_b \vert)^{1/2}$, 
which is essentially the same result obtained by equating the Gaussian and BKT critical
temperatures.

%
%
Vortex-antivortex pairs have been detected in ultracold atoms using absorption images 
of an expanding cloud~\cite{helmerson-2008}, and we expect that similar techniques can 
be used to detect vortex loops as the system evolves from anisotropic-3D to 2D regime.
Vortex loops should appear as dark rings in the image since there are no atoms to 
absorb light in their cores. At temperature $T$, the ratio of characteristic {\it in situ} 
core size of vortex loops in the $(x, y)$ plane $\xi_\perp (T) = \xi_{0,\perp} |\epsilon(T)|^{-2/3}$ 
and along the $\mathbf{z}$ direction $\xi_{z} (T) = \xi_{0,z} |\epsilon(T)|^{-2/3}$
is $\xi_{\perp} (T)/ \xi_z (T) \approx 0.91$ for $\eta = 2$ and $k_{2D} a_z = 2.2$.
Since typical values of $\xi_{0,\perp} \approx 0.5 {\rm \mu m}$, then
$\xi_{\perp} (T) \approx 2.3 {\rm \mu m}$ and $\xi_z (T) \approx 2.5 {\rm \mu m}$ 
at temperatures $T = 0.9 T_c$, and vortex loops extend to nearly six planes for 
an optical lattice with $a_z \approx 0.43 {\rm \mu m}$.
Smaller values of $\eta$ or larger values of $k_{2D} a_z$ enlarge $\xi_{z} (T)$.
For parameters $\eta = 1.7$, $k_{2D} a_z = 5.0$ and $T = 0.9 T_c$,  
the ratio $\xi_{\perp} (T)/ \xi_z (T) \approx 0.34$ and $\xi_z (T) \approx 6.8 {\rm \mu m}$, 
such that vortex loops extend to nearly sixteen planes in optical lattices with 
$a_z \approx 0.43 {\rm \mu m}$.

%
%
We analyzed the finite temperature phase diagram of attractive fermion mixtures in 1D 
optical lattices. At low temperatures, we found that a dimensional crossover from an 
anisotropic-3D (BCS) to an effectively 2D (BKT) superfluid occurs as a function 
of attraction strength eventhough the tunneling amplitude is fixed. In addition, we discussed 
that vortex excitations change from elliptical multiplane vortex loops in the anisotropic-3D regime to 
planar vortex-antivortex pairs in the 2D regime, and suggested an experiment to detect their presence. 

We thank the NSF (DMR-0709584) for support.

\end{document}